\documentclass[sn-mathphys,Numbered]{sn-jnl}

\usepackage{graphicx}%
\usepackage{multirow}%
\usepackage{amsmath,amssymb,amsfonts}%
\usepackage{amsthm}%
\usepackage{mathrsfs}%
\usepackage[title]{appendix}%
\usepackage{xcolor}%
\usepackage{textcomp}%
\usepackage{manyfoot}%
\usepackage{booktabs}%
\usepackage{algorithm}%
\usepackage{algorithmicx}%
\usepackage{algpseudocode}%
\usepackage{listings}%
\usepackage{adjustbox}
\usepackage{mathdots}

\theoremstyle{thmstyleone}%
\theoremstyle{thmstyletwo}%

\theoremstyle{thmstylethree}%

\begin{document}

\title[Article Title]{
  \begin{center}
    \textbf{Observation of broadband super-absorption\\ of electromagnetic waves \\ through space-time symmetry breaking}
  \end{center}
}

\author{\fnm{Matteo} \sur{Ciabattoni}}\email{mc2574@cornell.edu}
\equalcont{These authors contributed equally to this work.}

\author{\fnm{Zeki} \sur{Hayran}}\email{zh337@cornell.edu}
\equalcont{These authors contributed equally to this work.}

\author*{\fnm{Francesco} \sur{Monticone}}\email{francesco.monticone@cornell.edu}

\affil{School of Electrical and Computer Engineering, Cornell University, Ithaca, New York 14853, USA}

\abstract{Using time as an additional design parameter in electromagnetism, photonics, and wave physics is attracting considerable research interest, motivated by the possibility to explore physical phenomena and engineering opportunities beyond the physical limits of time-invariant systems. However, despite substantial theoretical promise, the practical realization and observation of many new effects and capabilities leveraging such temporal degrees of freedom have remained elusive. Here, we report the first experimental demonstration of enhanced broadband absorption of electromagnetic waves in a continuously modulated time-varying system, exceeding one of the key theoretical limits of linear time-invariant absorbers. This is achieved by harnessing the frequency-wavevector transitions and enhanced interference effects enabled by breaking both continuous space- and time-translation symmetries in a periodically time-modulated absorbing structure operating at radio-frequencies. Furthermore, we demonstrate broadband coherent wave absorption using a secondary control wave, observing a nearly perfect, reconfigurable, anti-reflection effect over a broad continuous bandwidth. Our findings provide new insights to challenge existing paradigms on the limits of wave absorption and may pave the way to the development of devices that operate in a regime fundamentally beyond the reach of any linear time-invariant system.}


\maketitle

\section{Introduction}\label{sec1}

\begin{figure}
 \centering
 \begin{adjustbox}{width=1.05\linewidth, center}
  \includegraphics[keepaspectratio]{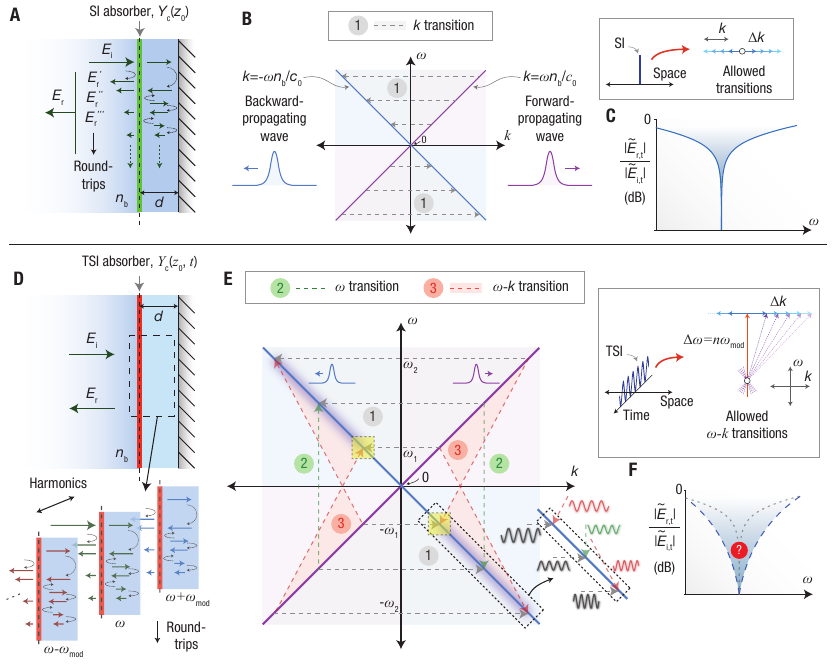}
  \end{adjustbox}
\caption{\textbf{Enhanced absorption using a periodically time-varying absorber.} \textbf{(A)} Schematic of a linear time-invariant absorber, based on a thin lossy layer acting as a spatial interface (SI) backed by a PEC sheet, separated by a spacer medium of thickness \( d \). The incoming normally incident wave undergoes multiple round-trips within the spacer medium, leading to interference between multiple reflected fields. \textbf{(B)} Dispersion diagram of the forward- (purple line) and backward- (blue line) propagating modes (plane waves) in a spatially-homogeneous background medium with refractive index \( n_{\text{b}} \), alongside the coupling/transitions induced by a SI. As a spatial boundary conserves the frequency of the incident wave, coupling occurs only between modes of identical frequency, but different (opposite) wavevectors. \textbf{(C)} Resulting ratio between the magnitude of the reflected field (\( E_{\text{r,t}} \)) and the magnitude of the incident field (\( E_{\text{i,t}} \)), typically optimized at a specific frequency, illustrating the limitations of linear time-invariant absorbers. \textbf{(D)} A linear time-varying absorber, based on a time-modulated thin lossy layer acting as a temporal-spatial interface (TSI), introducing not only multiple round trips within the spacer medium but also an additional degree of freedom: multiple harmonics that can couple to each other and enhance interference effects at different frequencies. \textbf{(E)} Dispersion diagram of the background-medium modes (plane waves) and the coupling/transitions induced by a TSI. As the TSI can now couple different frequencies \( \omega \) and wavevectors \( k \), a richer array of interaction effects emerges, as described in the main text. These complex interactions can potentially reduce reflections via the destructive interference resulting from multiple transitions (bottom-right inset of panel (E)), enabling performance beyond what is possible with conventional time-invariant systems \textbf{(F)}. For conciseness, in (E), only the transitions that couple the forward-propagating modes to the backward-propagating modes across an illustrative input bandwidth are shown. }
    \label{fig1}
\end{figure}

The absorption of electromagnetic energy underpins critical technologies ranging from energy harvesting to stealth. The demand for compact and versatile solutions in these fields drives the need for thin absorbers that can be integrated into various devices and systems without compromising their form or function \cite{ra2015thin}. However, achieving efficient absorption within a thin (possibly subwavelength) structure poses considerable challenges due to the weak interaction between electromagnetic waves and lossy materials over a small thickness \cite{lin201990}. A common strategy to enhance absorption relies on partial reflections/scattering at interfaces to make the propagating wave undergo multiple round trips, hence increasing the wave-material interaction (maximized near a resonance) and leveraging interference effects to minimize the total back-reflection \cite{kats2016optical}. Fig. \ref{fig1}(A) illustrates a simple implementation of this approach, featuring a thin lossy sheet backed by a perfect electric conductor (PEC), separated by a spacer that may either be filled with a specific material or remain as the background medium. The thin sheet itself, when considered in a homogeneous background medium, acts as a spatial interface that breaks continuous translation symmetry along a certain spatial direction. This induces reflections or, in more general terms, wavevector- (\( k \)-)transitions between forward- and backward-propagating modes of the background medium, i.e., plane waves (see Fig. \ref{fig1}(B)). When this thin lossy sheet is used in the configuration shown in Fig. \ref{fig1}(A), which is at the basis of traditional absorber designs like the Salisbury screen \cite{fante1988reflection}, the resulting multiple round trips of the electromagnetic wave can then reduce the total reflection (hence, achieving enhanced absorption) through the destructive interference of reflected fields from each round trip. This can also be interpreted, equivalently, as a form of impedance matching: the input impedance of the structure is designed to closely match the wave impedance of the background medium. Importantly, however, the phase accumulated during these round trips (and the input impedance) is markedly frequency-dependent, influenced by the thickness \( d \) and the material of the spacer medium. Hence, perfect destructive interference of the reflected fields can only occur at discrete frequencies, and an evident tradeoff emerges between (imperfect) reflection reduction and bandwidth, as illustrated in Fig. \ref{fig1}(C). Physical intuition also suggests that reducing reflection is more difficult for small thicknesses and longer wavelengths, as accumulating sufficient phase delay to achieve destructive interference becomes increasingly challenging.

In light of these considerations, one may wonder whether introducing additional degrees of freedom to control interference effects could potentially enhance the efficiency of absorbers. However, adding new \emph{spatial} degrees of freedom, for example introducing multiple layers, can help only to a certain extent, as quantified by well-established theoretical limits, such as the Rozanov bound \cite{rozanov2000ultimate}. This bound formalizes the physical intuition outlined above, defining a fundamental tradeoff between absorption bandwidth, reflection reduction, and thickness, for any linear time-invariant absorber (LTI) that is transversely invariant and backed by a mirror. Conversely, introducing new \emph{temporal} degrees of freedom, for example changing the properties of the lossy material in time, may offer a route to surpass these limitations, but has remained relatively unexplored until recently. In this context, recent advances in time-varying systems have indeed demonstrated the possibility of enhanced control over both the spatial and temporal profiles of waves \cite{caloz2019spacetime, galiffi2022photonics, hayran2022homega}, introducing a new paradigm of ``metamaterials with high degrees of freedom'' \cite{engheta2020metamaterials}. Moreover, a few recent studies have explored, theoretically, the use of time-varying components to overcome performance limitations associated with absorption \cite{li2021temporal, firestein2022absorption, hayran2024beyond,mostafa2022coherently}, impedance matching \cite{shlivinski2018beyond}, pulse delaying \cite{hayran2021capturing}, and antenna radiation \cite{li2019beyond} (which was later experimentally validated \cite{mekawy2021parametric}, followed by another experimental study \cite{mostafa2023antenna} based on a different concept), among others. For a comprehensive review, we refer the reader to Ref. \cite{hayran2023using}. We also note that two recent studies have experimentally demonstrated performance beyond the conventional limits of LTI systems in the contexts of electromagnetic \cite{yang2022broadband} and acoustic \cite{wang2024breaking} absorption. The method outlined in Ref. \cite{yang2022broadband} involves instantaneously switching the system state (closing the structure) immediately after an incident pulse has fully entered, requiring a structure thicker than the spatial width of the pulse itself,
whereas the technique described in Ref. \cite{wang2024breaking} for enhancing acoustic absorption, which relies on prior information about the incoming signal acquired through a microphone, is infeasible for electromagnetic applications, as it would require transmission of information at superluminal speeds. This underscores the need for an alternative general approach to enhance electromagnetic absorption beyond the limits of linear time-invariant systems and calls for experimental demonstrations to validate such method. Here, we address some of these issues by proposing and experimentally demonstrating an approach that builds upon the theoretical works in Refs. \cite{mostafa2022coherently, hayran2024beyond} to boost electromagnetic absorption by leveraging temporal degrees of freedom. 
%
In particular, Ref. \cite{mostafa2022coherently} theoretically explored absorption enhancements in a time-varying structure in a configuration similar to ours in Fig. \ref{fig1}, showing that multiple reflection zeros can be induced via coherent absorption at distinct isolated frequencies using a separate monochromatic control wave. While multi-frequency perfect absorption in a ultra-thin structure, as theoretically demonstrated in \cite{mostafa2022coherently}, is indeed impossible to achieve in passive LTI systems, Ref. \cite{mostafa2022coherently} does not directly compare the resulting absorption enhancement against the Rozanov bound and does not explicitly address the possibility of suppressing reflections \textit{over a continuous bandwidth}, since an additional monochromatic incident wave with a frequency that lies between the generated reflection zeros will not necessarily experience absorption enhancement unless an additional temporal modulation component with a suitable modulation frequency is added to the original modulation. Furthermore, it remains unclear whether substantial absorption enhancements can be achieved across a broad frequency range when strong dispersion effects on the generated harmonics are considered. In contrast, here we experimentally demonstrate that, rather surprisingly, a single periodic modulation of the absorbing element, without a separate control signal, can achieve absorption enhancement quantitatively breaking the Rozanov bound even for a continuously broadband incident signal with a large fractional bandwidth. Specifically, we will show that the \textit{broadband} incident wave itself serves as a ``control'' wave, enhancing its own absorption, through interference between harmonics, across its entire continuous bandwidth. To the best of our knowledge, these results represent the first experimental demonstration of a continuously modulated time-varying system that surpasses the theoretical limitations of linear time-invariant electromagnetic absorbers.

\section{Main}\label{sec2}

While the absorber configuration in Fig. \ref{fig1}(A) is based on an interface that breaks space-translation symmetry, we consider here a similar configuration but with an interface that also breaks continuous translation symmetry in time, inducing scattering in both wavevector and frequency domain, as illustrated in Fig. \ref{fig1}(D,E). This space-time interface consists of a thin lossy layer positioned at \( z_0 \), modeled as a non-dispersive time-varying sheet admittance \( Y_{\text{c}} \), which relates the surface current density to the electric field along the sheet \cite{peuzin1994demonstration} (for a thin layer, \( Y_{\text{c}} \) is simply given by the conductivity $\sigma$ of the material times the thickness $t$ \cite{tretyakov2003analytical}). The sheet is subject to a periodic temporal modulation, expressed as \( Y_{\text{c}}(z_0,t) = Y_{\text{c}}^0[1 + M_{\text{mod}} f(t,\omega_{\text{mod}}, \varphi_{\text{mod}})] \), where \( M_{\text{mod}} \) is the modulation amplitude and \( f(.) \) is a time-periodic function with a modulation frequency \( \omega_{\text{mod}} \) and phase \( \varphi_{\text{mod}} \). As depicted in Fig. \ref{fig1}(D), a periodic modulation generates additional harmonics \( \omega + n \omega_{\text{mod}} \) (where \( n \in \mathbb{Z} \) is the harmonic number), which are coupled to each other and to the fundamental frequency \( \omega \), hence introducing what is often termed a ``synthetic dimension'' in the photonics literature \cite{yuan2018synthetic, ehrhardt2023perspective}. Consequently, the interference of waves in this configuration is determined not only by multiple round trips in space but also through these additional harmonics, and the amplitude and phase of the reflected waves for these harmonics can be adjusted using the modulation settings, such as \( M_{\text{mod}} \) and \( \varphi_{\text{mod}} \) \cite{zang2019nonreciprocal}. The richer wave dynamics arising from such temporal and spatial scattering will be harnessed to suppress reflections, and enhance absorption, beyond what would be possible with spatial scattering only. Throughout this article, a thin absorber with dynamically modulated admittance \( Y_{\text{c}}(z_0,t) \) will be termed as a time-space-interface (TSI) absorber, reflecting its role as a boundary that modulates wave behavior across both space and time. Conversely, a conventional time-invariant thin absorber will be termed a spatial-interface (SI) absorber.

As illustrated in Fig. \ref{fig1}(E), the TSI absorber is capable of inducing transitions both in frequency (\( \omega \)) and frequency-wavevector (\( \omega \)-\( k \)) domain, influenced by the bandwidth of the input wave and the modulation frequency. A \( \omega \)-only transition occurs when \( \omega = \omega_{\text{mod}}/2 \), allowing a positive frequency to couple with its negative counterpart, with the same wavevector/momentum \cite{landau2013mechanics,collin2001foundations} (Fig. \ref{fig1}(E), dashed green arrows). Conversely, a \( \omega \)-\( k \) transition arises between different frequencies 
with a different wavevector due to the dispersion relation in the considered structure (Fig. \ref{fig1}(E), dashed red arrows). Importantly, such \( \omega \)-\( k \) transitions are absent in time-varying systems that are spatially invariant (as in the well-studied case of wave propagation in a spatially homogeneous time-varying medium \cite{asgari2024photonic}) owing to wavevector conservation in a spatially-invariant system and the resulting wavevector mismatch between different frequency components. Instead, in a TSI absorber, the spatially-localized time-varying profile of the absorbing sheet does not conserve wavevector and can ensure the necessary momentum matching to induce such \( \omega \)-\( k \) transitions. Note that, due to the reality of electromagnetic fields, the quadrant in the dispersion diagram where \(\omega > 0\) and \(k > 0\) (\(\omega < 0\) and \(k > 0\)) is equivalent to the quadrant where \(\omega < 0\) and \(k < 0\) (\(\omega > 0\) and \(k < 0\)) because they describe the same physical oscillation \cite{pendharker2017pt}. Thus, the \( \omega \)-only and \( \omega \)-\( k \) transitions numbered as (2) and (3) in Fig. \ref{fig1}(E) may lead to reflected waves capable of destructively interfering with the waves resulting from the \(k\)-only transitions numbered as (1). Through a suitable selection of the modulation parameters, it may be possible to optimize these interactions over a broad bandwidth, thereby enhancing destructive interference between reflected fields, at different frequencies, from each round trip and harmonic component (Fig. \ref{fig1}(E), bottom-right inset). This can then potentially lead to a marked reduction in reflection, and an increase in absorption, of the incoming wave, as illustratively shown in Fig. \ref{fig1}(F), surpassing the performance of any LTI absorber having the same thickness.

To quantify these predictions and design the modulation parameters to maximize absorption over a broad bandwidth, we theoretically studied the TSI absorber structure in Fig. \ref{fig1}(D) using Floquet analysis \cite{galiffi2022photonics} (details in Supplementary Section 1), which provides an exact theoretical model to calculate the harmonics of the reflected field from the incident field, for any periodic modulation, accounting for all the interactions described above (unlike the simplified models used, for example, in our earlier theoretical work \cite{hayran2024beyond}, which only considered $\omega$-transitions). This approach yields the following equation that connects the incident electric field \( \Tilde{E}_{\text{i}}(\omega) \) to the reflected field \( \Tilde{E}_{\text{r}}(\omega) \) (throughout this article, quantities marked with a ``\textasciitilde'' denote variables in the frequency domain, specifically those dependent on the frequency of the incident field):

\begin{equation}
    \begin{bmatrix}
    \Tilde{\Psi}^{-n'} & \cdots & \cdots & Y_{\text{c}}^{-n'} & \cdots & \cdots & Y_{\text{c}}^{-2n'} \\
    \vdots & \ddots & \vdots & \vdots & \vdots & \iddots & \vdots \\
    \vdots & \cdots & \Tilde{\Psi}^{-1} & Y_{\text{c}}^{-1} & Y_{\text{c}}^{-2} & \cdots & \vdots \\
    Y_{\text{c}}^{n'} & \cdots & Y_{\text{c}}^{1} & \Tilde{\Psi}^{0} & Y_{\text{c}}^{-1} & \cdots & Y_{\text{c}}^{-n'} \\
    \vdots & \cdots & Y_{\text{c}}^{2} & Y_{\text{c}}^{1} & \Tilde{\Psi}^{1} & \cdots & \vdots \\
    \vdots & \iddots & \vdots & \vdots & \vdots & \ddots & \vdots \\
    Y_{\text{c}}^{2n'} & \cdots & \cdots & Y_{\text{c}}^{n'} & \cdots & \cdots & \Tilde{\Psi}^{n'}
    \end{bmatrix}
    \begin{bmatrix}
    \Tilde{E}_{\text{r}}^{-n'} \\
    \vdots \\
    \Tilde{E}_{\text{r}}^{-1} \\
    \Tilde{E}_{\text{r}}^0 \\
    \Tilde{E}_{\text{r}}^1 \\
    \vdots \\
    \Tilde{E}_{\text{r}}^{n'}
    \end{bmatrix}
    =
    \begin{bmatrix}
    -Y_{\text{c}}^{-n'} \\
    \vdots \\
    -Y_{\text{c}}^{-1} \\
    Y_{\text{a}} - Y_{\text{c}}^{0} - \Tilde{Y}_{\text{in}}^{0} \\
    -Y_{\text{c}}^{1} \\
    \vdots \\
    -Y_{\text{c}}^{n'}
    \end{bmatrix} \Tilde{E}_{\text{i}},
    \label{eq1}
\end{equation} 

\noindent where

\begin{equation}
    \Tilde{\Psi}^{n} = Y_{\text{a}} + \Tilde{Y}_{\text{in}}^{n} + Y_{\text{c}}^0.
    \label{eq2}
\end{equation}

Here \( \Tilde{E}_{\text{r}}^{n} \) denotes the \( n \)th harmonic of the reflected field at \( \omega + n \omega_{\text{mod}} \) (with \( n \) being an integer running from \( -n' \) to \( n' \), where \( n' \) is the maximum considered harmonic number). \( Y_{\text{a}} \) represents the wave admittance of the non-dispersive background medium of incidence and \( \Tilde{Y}_{\text{in}} \) denotes the input admittance of the metal-backed spacer medium (after the absorbing sheet), as further discussed in Supplementary Section S1. 

To gain physical insight into the effect of the modulation parameters on the reflection harmonics, we can consider a simple ``two-level'' system where the only relevant reflected field harmonics are \( \Tilde{E}_{\text{r}}^{0} \) and \( \Tilde{E}_{\text{r}}^{-1} \). Assuming a temporal modulation in the form of \( Y_{\text{c}}(z_0,t) = Y_{\text{c}}^0[1 + M_{\text{mod}} \cos(\omega_{\text{mod}}t + \varphi_{\text{mod}})] \) and employing the rotating-wave approximation \cite{allen1987optical}, we obtain through Eq. (\ref{eq1}) the relation between \( \Tilde{E}_{\text{r}}^{-1} \) and the incident field \( \Tilde{E}_{\text{i}} \) as

\begin{equation}
    \Tilde{E}_{\text{r}}^{-1} \approx \frac{ \pi Y_{ \text{a}} Y_{\text{c}}^0 M_{\text{mod}} e^{-i \varphi_{\text{mod}}}}{\Tilde{\Psi}^{0} \Tilde{\Psi}^{-1}} \Tilde{E}_{\text{i}},
\end{equation}

\noindent which shows that the amplitude and phase of the reflected harmonic can be controlled directly through the modulation parameters \( M_{\text{mod}} \) and \( \varphi_{\text{mod}} \). As shown in the following, this added flexibility in controlling reflections afforded by a temporal modulation can be harnessed to enhance absorption performance.

While the previous discussion focused on monochromatic incident fields, we now consider the case of a multi-frequency excitation. The \( n \)th harmonic of the reflected field, \( \Tilde{E}_{\text{r}}^{n}\), for an incident field \( \Tilde{E}_{\text{i}} \) at frequency \( \omega' \), can be conveniently expressed using a two-frequency function as \( \Tilde{E}_{\text{r}}(\omega' + n \omega_{\text{mod}}, \omega') \). Then, for a multi-frequency incident field, the total reflected field \( \Tilde{E}_{\text{r,t}} \) can be obtained by integrating over \( \omega' \),

\begin{equation}
    \Tilde{E}_{\text{r,t}}(\omega) = \int_{-\infty}^{\infty} \Tilde{E}_{\text{r}}(\omega, \omega') \, d\omega'.
    \label{eq3}
\end{equation}

\noindent Consider the simple case of an input field with two monochromatic components, \( E_{\text{i,t}} = \text{A}_1 \cos{(\omega_1 t)}  + \text{A}_2 \cos{(\omega_2 t)} \), where \( \text{A}_1 \) and \( \text{A}_2 \) are real coefficients. With a periodic time modulation at frequency \( \omega_{\text{mod}} = \omega_2 - \omega_1 \) (\( \omega_1 < \omega_2 \)), the reflected field at frequency \( \omega_1 \) comprises two components: (i) \( \Tilde{E}_{\text{r}}(\omega_1, \omega_1) \) + \( \Tilde{E}_{\text{r}}(-\omega_1, -\omega_1) \), i.e., the reflection originating from the fundamental (\( n = 0 \)) harmonic of the incident field component of frequency \( \omega_1 \) [corresponding to horizontal transitions denoted as (1) in Fig. \ref{fig1}(E)], and (ii) \( \Tilde{E}_{\text{r}}(-\omega_1, \omega_2) + \Tilde{E}_{\text{r}}(\omega_1, -\omega_2) \), i.e., the reflection originating from the \( n = \pm1 \) harmonic of the incident field component of frequency \( \omega_2 \) [corresponding to oblique transitions denoted as (3) in Fig. \ref{fig1}(E)]. Thus, the total reflected field at \( \omega_1 \) becomes \( \Tilde{E}_{\text{r,t}}(\pm\omega_1) = \Tilde{E}_{\text{r}}(\pm\omega_1, \pm\omega_1) + \Tilde{E}_{\text{r}}(\pm\omega_1, \mp\omega_2) \) (ignoring higher-order harmonics), resulting in what we term ``non-degenerate interference,'' namely, interference, at the same frequency \( \omega_1 \), between reflected fields originating from different frequency components of the incident field, \( \omega_1 \) and \( \omega_2 \) (corresponding to the marked yellow squares in Fig. \ref{fig1}(E)). As discussed above, the amplitude and phase of the reflection harmonics can be adjusted through the modulation parameters \( M_{\text{mod}} \) and \( \varphi_{\text{mod}} \). Consequently, by appropriately selecting these parameters, one can ensure complete destructive interference between \( \Tilde{E}_{\text{r}}(\omega_1, \omega_1) \) and \( \Tilde{E}_{\text{r}}(\omega_1, \omega_2) \) to reduce the total reflected field \( \Tilde{E}_{\text{r,t}}(\omega_1) \). 
Rather remarkably, as we will show in the following, this approach is effective even for continuously broadband input fields, despite the number of modulation parameters remaining the same. This can be understood by noting that if the LTI system is designed to operate near its fundamental limit, any additional ``opportunity'' for reflection suppression, enabled by the described non-degenerate interference effects, may be sufficient to enhance performance beyond the LTI limit, for any input field within a certain bandwidth.

\begin{figure}
 \centering
  \includegraphics[width=1.0\linewidth, keepaspectratio]{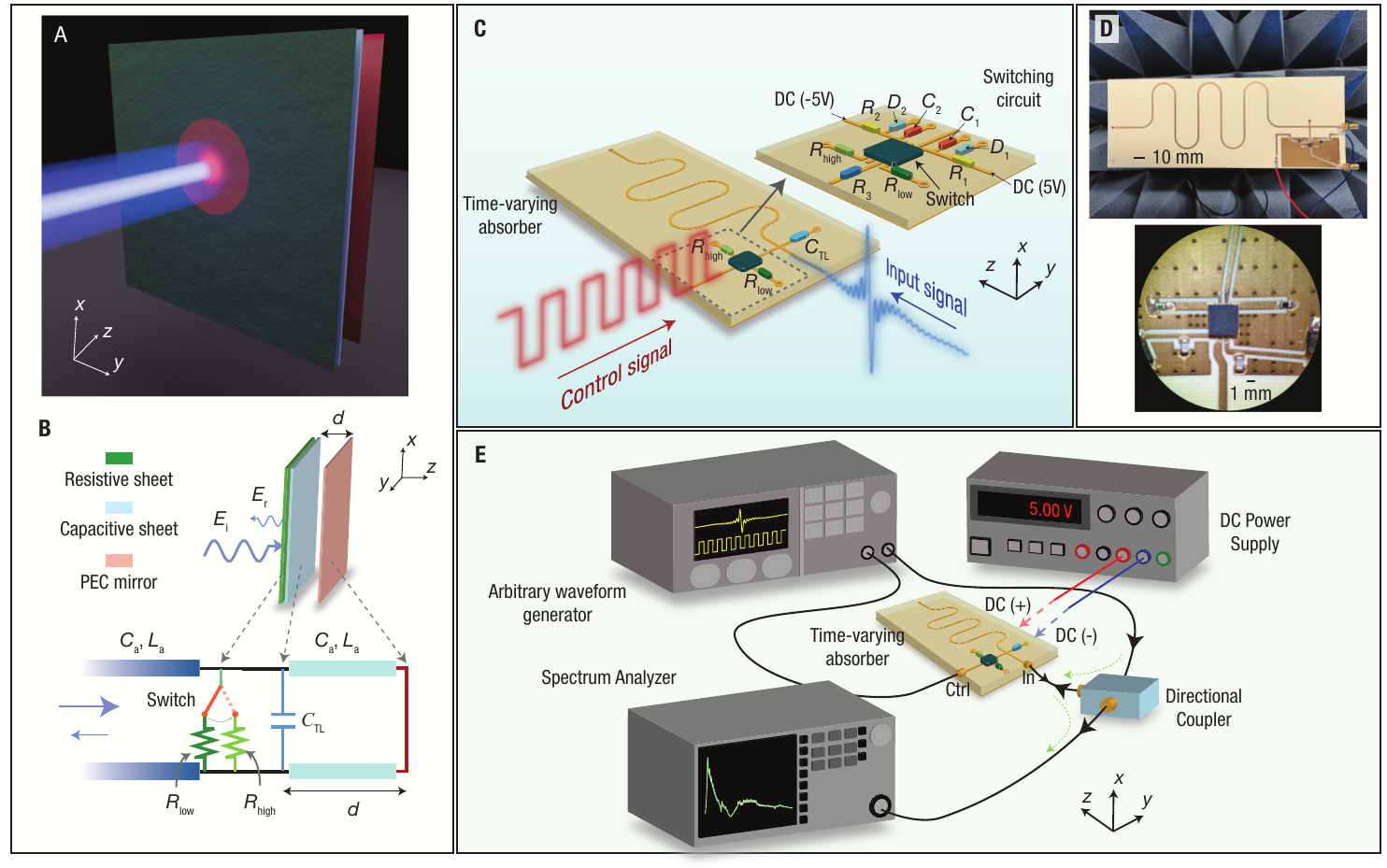}
    \caption{\textbf{Experimental realization of the time-varying absorber.} \textbf{(A)} Illustration of a TSI absorber composed of thin sheet with both conductive and dielectric properties (e.g., a lossy dielectric sheet) backed by a perfect electric conductor (PEC), with an air spacer medium in between. The surface admittance of the sheet is modulated over time. \textbf{(B)} The structure in (A) is modeled (assuming normal wave incidence) and implemented in the form of a transmission-line (TL) circuit, where the dielectric and conductive properties of the sheet are realized as a capacitor and a time-varying resistor, respectively, connected in parallel. The resistor value is dynamically altered by a RF switch that toggles between two predefined resistances. The air spacer backed by a PEC plane is mimicked through the use of a shorted transmission line. \textbf{(C)} Illustration and  \textbf{(D)} photograph of the TL-based absorber and its circuit elements on a printed circuit board (PCB). The meandering configuration of the microstrip line is intended solely to utilize the available space efficiently and does not influence the absorption characteristics of the absorber. The lower inset in (D) displays a microscope image of the RF switch. \textbf{(e)} Measurement setup, which includes an arbitrary waveform generator (AWG) to generate both the control signal for the switch and the input signal, a DC supply to bias the switch, and a spectrum analyzer to measure the reflected spectrum.}
    \label{fig2}
\end{figure}

As a proof-of-concept experimental demonstration of enhanced absorption performance, we implemented the proposed time-varying absorber in the form of a time-varying transmission-line circuit operating at radio-frequencies (RF), as depicted in Figs. \ref{fig2}(A) and \ref{fig2}(B). In our design, the time-modulated absorbing sheet is implemented using an RF switch that periodically alternates between two resistors, \( R_{\text{low}} \) and \( R_{\text{high}} \), at a modulation frequency \( \omega_{\text{mod}} \). A dielectric sheet, realized as a shunt capacitance in the transmission-line system, was also introduced at \( z = z_0 \) to optimize the absorption performance and approach the theoretical absorption limit in the LTI case. Both the schematic and the photograph of the constructed transmission line-based TSI absorber are shown in Figs. \ref{fig2}(C) and \ref{fig2}(D), respectively, while the measurement setup is illustrated in Fig. \ref{fig2}(E). For additional details about the experimental fabrication and measurements, please refer to the Methods section.

Having established both a theoretical framework and an experimental platform, we now proceed to assess the absorption performance of the proposed time-varying absorber. Here, the figure of merit for absorption efficiency, denoted as \( \eta_{\text{abs}} \), is defined as follows:

\begin{equation}
    \eta_{\text{abs}} = 1 - \frac{U_{\text{ref}}}{U_{\text{in}}} = 1 - \frac{\int_{-\infty}^{\infty} |\Tilde{E}_{\text{r,t}}(\omega)|^2 d\omega}{\int_{-\infty}^{\infty} |\Tilde{E}_{\text{i,t}}(\omega)|^2 d\omega},
    \label{eq4}
\end{equation} 

\noindent where \( U_{\text{ref}} \) and \( U_{\text{in}} \) are the total reflected and incident wave energy densities, respectively. Since our goal is to experimentally demonstrate that a time-varying absorber can perform better than even its best time-invariant counterpart, we first need to establish a fundamental upper bound for the figure of merit \( \eta_{\text{abs}} \) for LTI absorbers. To do so, we consider the Rozanov limit mentioned above \cite{rozanov2000ultimate}, which can be written in the form of an integral inequality for the reflection coefficient \( \Gamma \):

\begin{equation}
    -\int_{0}^{\infty} \ln(|\Gamma(\lambda)|) \, d\lambda \leq 2 \pi^2 d \mu_{\text{s,b}} \sqrt{\frac{\varepsilon_{\text{s,a}}}{\mu_{\text{s,a}}}}.
    \label{eq5}
\end{equation} 

\noindent Here, the Rozanov bound was generalized for an incident medium with arbitrary permittivity and permeability, expanding upon the original derivation in Ref. \cite{rozanov2000ultimate}, which assumed free space as the incident medium (see Supplementary Section S2). In Eq. (\ref{eq5}), \( \lambda \) represents the free space wavelength, \( d \) is the total thickness of the structure, \( \mu_{\text{s,b}} \) is the static relative magnetic permeability of the spacer medium, and \( \varepsilon_{s,a} \) and \( \mu_{\text{s,a}} \) are the static relative electric permittivity and magnetic permeability of the incident medium, respectively. For conciseness, in the following we denote the right-hand side of Eq. (\ref{eq5}) as \( I_{\text{R}} \). For the transmission-line implementation of the absorber, \( I_{\text{R}} \) can be found as \( I_{\text{R}} = 2 \pi^2 d c_0 L_{\text{s,b}} \sqrt{C_{\text{s,a}}/L_{\text{s,a}}} \) using the standard equivalence principle between plane-wave propagation in layered optical material systems and voltage/current-wave propagation in transmission-line systems \cite{caloz2005electromagnetic} (see Supplementary Section S3 for details). Here, \( c_0 \) is the speed of light, \( L_{\text{s,b}} \) represents the static inductance per unit length of the spacer transmission line, while \( L_{\text{s,a}} \) and \( C_{\text{s,a}} \) denote the static inductance and capacitance per unit length of the incident transmission line, respectively. 

Finding the precise upper bound for \( \eta_{\text{abs}} \) from Eq. (\ref{eq5}) is not trivial, since Eq. (\ref{eq4}) involves integration in frequency whereas Eq. (\ref{eq5}) involves integration in wavelength. However, employing variational calculus (see Supplementary Section S2 for details) we derived the following expression for the maximum possible \( \eta_{\text{abs}} \):

\begin{equation}
    \eta_{\text{abs,R}} = 1 - \frac{\psi}{2 \omega_1^* \omega_2} - \frac{\omega_1^* - \omega_1}{\omega_2 - \omega_1}\quad,
    \label{eq6}
\end{equation} 

\noindent where \( \omega_1^* \) is defined in Supplementary Section S2, and

\begin{equation}
    \psi = 2 \exp \left\{ 2 \left( -\frac{I_\mathrm{R}}{2\pi c_0} + \frac{\ln(\omega_1^*) + 1}{\omega_1^*} - \frac{\ln(\omega_2) + 1}{\omega_2} \right) / \left[ \frac{1}{\omega_1^*} - \frac{1}{\omega_2} \right] \right\},
    \label{eq7}
\end{equation} 

\noindent for an incident field with a normalized spectrum defined as

\begin{equation}
    |\Tilde{E}_{\text{i,t}}(\omega)| = 
    \begin{cases} 
    1 & \text{for } \omega_1 \leq |\omega| \leq \omega_2 \\
    0 & \text{otherwise}.
    \end{cases}
    \label{eq8}
\end{equation} 

\noindent Equations (\ref{eq6}) and (\ref{eq7}) establish the upper limit for the chosen figure of merit, \( \eta_{\text{abs}} \), for any passive LTI absorber that is transversely invariant and is backed by a perfect mirror (for normal incidence, as in \cite{rozanov2000ultimate}, but these results can also be generalized to the oblique incidence case). Our analytic derivation of this bound was also independently validated against results obtained through a genetic optimization algorithm, demonstrating excellent agreement between these methodologically distinct approaches to identify optimal absorption performance (see Supplementary Section S4). We note that an analytic bound offers substantial advantages over numerical methods, as it avoids the extensive computational demands of optimization techniques and indicates the limits of what is possible from the start.

\begin{figure}
 \centering
 \begin{adjustbox}{width=1.3\linewidth, center}
  \includegraphics[keepaspectratio]{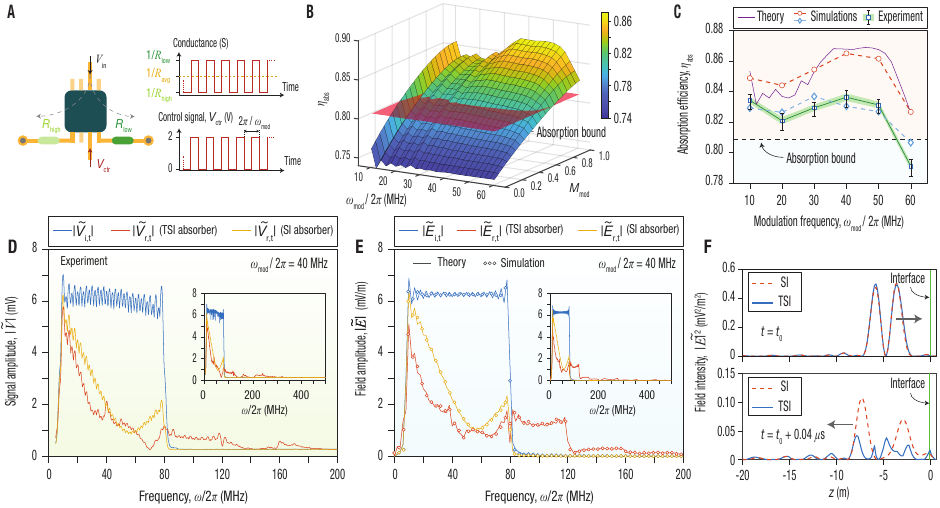}
  \end{adjustbox}
    \caption{\textbf{Experimental demonstration of absorption performance beyond the limits of LTI systems.} \textbf{(A)} Illustration of an RF switch that alternates between two resistor values, \( R_{\text{low}} \) and \( R_{\text{high}} \), controlled by a periodically varying signal \( V_{\text{ctr}} \), with a modulation angular frequency \( \omega_{\text{mod}} \). \textbf{(B)} Theoretical calculations (obtained through Eq. (\ref{eq1}) and (\ref{eq4})) showing that an increase in the modulation amplitude \( M_{\text{mod}} \) enhances the absorption efficiency \( \eta_{\text{abs}} \), surpassing the fundamental absorption limit of linear time-invariant (LTI) systems across a range of modulation frequencies \( \omega_{\text{mod}} \). The considered incident field has a very broad bandwidth of 72 MHz around a center frequency of 44 MHz. \textbf{(C)} Superimposed theoretical, simulation, and experimental results for \( M_{\text{mod}} = 0.99 \), demonstrating that the TSI absorber exceeds the LTI absorption bound. Orange circles are computed with Ansys Lumerical \cite{lumerical} for an ideal time-varying absorber, while blue diamonds are computed with CST Microwave Studio 2022 \cite{CST} for the circuit-equivalent absorber. \textbf{(D, E)} Detailed spectra of the incident and reflected fields for \( \omega_{\text{mod}} / 2\pi = 40 \) MHz for the experimental (d) and theoretical and simulation cases (E), showing substantial reflection reduction over a broad frequency range in the time-varying case. \textbf{(F)} Simulated time snapshots of the incident and reflected field intensity envelopes before (at time \(t = t_{o}\), upper panel) and after (at time \(t = t_{o} + 0.04\mu\text{s}\), lower panel) interaction with the SI (red dashed line) and TSI (blue solid line) absorbers. Additional details and all values of the considered parameters are provided in Supplementary Section S6.}
    \label{fig3}
\end{figure}

Employing the theoretical and experimental framework detailed above, our results are summarized in Fig. \ref{fig3}. For the theoretical calculations, the time modulation is a rectangular function that oscillates between \( R_{\text{low}} \) and \( R_{\text{high}} \), as illustrated in Fig. \ref{fig3}(A). The computed absorption efficiency, \( \eta_{\text{abs}} \), calculated through Eqs. (\ref{eq1}) and (\ref{eq4}) (with \( n' = 50 \)), is displayed for various \( \omega_{\text{mod}} \) and \( M_{\text{mod}} \) values in Fig. \ref{fig3}(B), where the optimal \( \varphi_{\text{mod}} \) is determined through a parameter sweep, for an incident field as in Eq. (\ref{eq8}) with a very broad bandwidth of 72 MHz around a center operational frequency of 44 MHz. These findings reveal that an increase in \( M_{\text{mod}} \) markedly enhances the absorption performance, surpassing the absorption bound \( \eta_{\text{abs,R}} \) of LTI absorbers (defined by Eq. (\ref{eq6})) for large enough modulation amplitudes. These theoretical results confirm that, even for a continuously broadband input field, adjusting the few available modulation parameters is sufficient to obtain absorption performance exceeding the LTI limit. Theoretical results are then compared with experimental measurements in Fig. \ref{fig3}(C) for \( M_{\text{mod}} = 0.99 \) (implemented using commercially available resistors, as further discussed in the Methods section). Error bars, indicative of experimental variations, are included, with methodological details provided in the Methods section. Notably, the experimental absorption efficiency \( \eta_{\text{abs}} \) (defined by Eq. (\ref{eq4})) surpasses the absorption bound \( \eta_{\text{abs,R}} \) (defined by Eq. (\ref{eq6})), following the same trend as the theoretically calculated efficiency, as expected. To independently verify these results, we also included in Fig. \ref{fig3} numerical full-wave simulations conducted using the finite-difference time-domain method (FDTD) and the finite integration technique, implemented in two distinct commercial software packages \cite{lumerical,CST}. For the first set of simulation results (orange circles), computed with Ansys Lumerical \cite{lumerical}, we simulated an ideal absorber structure, demonstrating good agreement with our theoretical calculations (purple line); the second set of results (blue diamonds) was instead computed using CST Microwave Studio 2022 \cite{CST} which, through its EM-Circuit Transient Co-Simulation tool, captures the main non-idealities of the circuit implementation, from the losses in the substrate to the non-idealities of the lumped-element components. As can be seen from \ref{fig3}(C), these simulation results show remarkable agreement with our experimental measurements. Notably, the absorption efficiency depends on the modulation frequency and decreases for modulation frequencies greater than 50 MHz. This reduction can be attributed to the more limited interaction between frequency components as the modulation-induced frequency transitions start exceeding the bandwidth of the incident signal, which results in fewer coupled harmonic components available to enhance absorption through interference. The remaining discrepancy between simulation and experimental results can be explained by the non-ideal behavior of the switch modulation function (characterized by finite rise and fall times). For more information on simulation methods, please refer to Supplementary Section S5. These results represent an important experimental step towards the long-sought-after goal of electromagnetic absorbers operating beyond the fundamental limits of LTI systems.

To provide additional insight into these results, Figs. \ref{fig3}(D) and \ref{fig3}(E) display detailed spectra for the incident and reflected fields for our theoretical, FDTD simulation, and experimental analyses in both the time-invariant (SI absorber) and time-varying (TSI absorber) scenarios. These findings confirm strong consistency between theoretical calculations, simulations, and experimental data, with the TSI configuration showing markedly lower reflection levels across a broad frequency spectrum compared to the SI configuration. For the TSI absorber, we observe additional harmonics extending beyond the original input bandwidth, an expected result of periodic time modulation. Importantly, however, the considered figure of merit, \( \eta_{\text{abs}} \), takes into account the reflected energy over the entire spectrum, including all harmonic contributions beyond the input signal bandwidth. Absorption markedly surpasses the established bounds for LTI absorbers despite the presence of these harmonics, thereby violating the absorption bound given by Eq. (\ref{eq6}), which in turn implies a violation of the bound given by Eq. (\ref{eq5}). Hence, our approach is notably different from other methods, such as spectrum spreading based on periodic or pseudorandom modulations \cite{chambers2005smart, wang2020spread}, which reduce reflection within a certain frequency range by spreading it over a broader range, but do not necessarily enhance absorption across the entire frequency spectrum. Moreover, Fig. \ref{fig3}(F) provides simulated temporal snapshots of the propagating wave pulse before and after interaction with the TSI absorber, clearly showing a strong reduction in the reflected field amplitude, in stark contrast with the case of a time-invariant SI absorber. The complete time-domain animation is provided as Supplementary Movie 1. Importantly, these results also confirm that the absorber structure can be much thinner than the spatial width of the incident pulse, in contrast with other methods based on time-switching, where the pulse needs to be fully contained in the absorber when the switching occurs \cite{yang2022broadband, firestein2022absorption}. Finally, we underscore that here the temporal modulation is restricted to the resistive component of the absorber, which ensures that no energy is exchanged with the modulating system \cite{mostafa2022coherently}. This is in contrast to most other time-varying systems studied in the literature, in which the temporal modulation of reactive components implies energy transfer with the modulating circuit \cite{caloz2019spacetime}. As a result, in our case all energy dissipation originates from the inherent ohmic losses of the resistors, with the temporal modulation serving to enhance the power dissipated in the resistive element, due to interference between harmonics, without any energy being extracted by the external modulator \cite{mostafa2022coherently}. For additional details, demonstrating that all absorbed energy (net energy input) is indeed dissipated in the time-varying resistive element, see Supplementary Section S7.

Finally, to further explore the potential of the enhanced interference effects enabled by a TSI absorber, we investigate coherent absorption effects involving a primary input wave and a secondary control wave, similar to the approach theoretically demonstrated in Ref. \cite{mostafa2022coherently}, but for continuously broadband signals. Traditional coherent wave absorbers typically require two waves at the same frequency coming from opposite directions to achieve perfect absorption at distinct, isolated frequencies \cite{chong2010coherent, baranov2017coherent}. In such a configuration, the phase of the secondary wave controls the total reflection of the system through interference between partially reflected and transmitted fields at the desired frequency. In our configuration, instead, we consider incoming waves from only one side, with different frequencies, and we harness the frequency coupling/transitions enabled by the temporal modulation to induce and engineer the interference between reflected fields over a range of frequencies. In this way, the coherent coupling between a primary input wave (Beam-1) and a secondary control wave (Beam-2) could be used to control, all-optically, the absorption and anti-reflection response of the structure, with the goal, for example, of eliminating all reflections across the primary wave bandwidth. 

\begin{figure}
 \centering
  \includegraphics[width=0.6\linewidth, keepaspectratio]{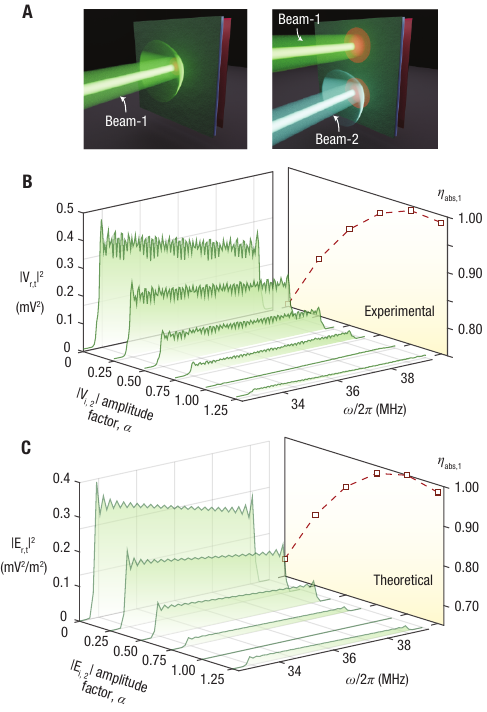}
    \caption{\textbf{Broadband coherent wave absorption using a time-varying absorber.} \textbf{(A)} Schematic illustration of the TSI-based coherent wave absorber, where a control beam (Beam-2) can modulate the reflection from a primary input beam (Beam-1), potentially eliminating it entirely over a continuous range of frequencies. \textbf{(B, C)} Experimental (B) and theoretical (C) results showing the spectra of the reflected field/voltage as a function of the Beam-2 amplitude factor \( \alpha \) (defined below), demonstrating that reflections can be actively controlled and even entirely suppressed by adjusting Beam-2. For clarity, Beam-1 and Beam-2 are spectrally well-separated, and the experimental and theoretical spectra only show the reflection response over the bandwidth of Beam-1. In (B) and (C) \( \eta_{\text{abs,1}} \) denotes the absorption efficiency calculated over the bandwidth of Beam-1. The amplitude factor \( \alpha \) is defined as a scaling factor applied to the specific amplitude profile of Beam-2 that achieves complete suppression of back-reflections across the bandwidth of Beam-1. The results in (B) are obtained using the circuit shown in Fig. \ref{fig2}(C). For details, see Supplementary Section S8.}
    \label{fig4}
\end{figure}

This mechanism is visually represented in Fig. \ref{fig4}(A), where the suppression of reflections from Beam-1 is achieved through Beam-2 at a different frequency range. The specific spectral profile of Beam-2 leading to the complete elimination of reflections for Beam-1 can be calculated using Eq. (\ref{eq1}) (for details, see Supplementary Section S7). Figs. \ref{fig4}(B) and (C) show both experimental and theoretical results, confirming our predictions. Rather remarkably, virtually zero reflection can be achieved across the entire bandwidth of interest, corresponding to unity absorption efficiency for Beam-1. We stress that this is impossible to achieve in any single-port LTI system, where zero reflection can only be obtained at isolated frequencies (i.e., over a zero-measure bandwidth), as predicted by Eq. (\ref{eq5}). This constraint is overcome in our time-varying system, demonstrating virtually perfect absorption over a continuous bandwidth for Beam-1. (It is important to note that we did not explicitly consider reflection reduction at Beam-2 frequencies, as the current absorber design is optimized specifically for Beam-1.)
This technique of ``linear control of light with light'' \cite{baranov2017coherent}, extended to time-varying systems and broadband operation (also recently studied in Ref. \cite{galiffi2023broadband} with individual time interfaces and in Ref. \cite{mostafa2022coherently} with periodic temporal modulations and multi-frequency incident signals) may enable the dynamic reconfiguration of a device response and the wideband modulation of waves without the need to alter the material properties or structural design of the system.

\section{Conclusion}\label{sec3}

In conclusion, we have reported the experimental demonstration of electromagnetic absorption performance beyond the fundamental limitations of linear time-invariant systems. To this end, we have taken inspiration from the basic idea of reducing reflections and enhancing absorption through destructive interference between partially reflected waves, extending it to time-varying systems. Based on this concept, we have designed and demonstrated a time-varying absorber, in the form of a periodically modulated absorbing sheet backed by a reflector, that takes advantage of the frequency-wavevector transitions and enhanced interference effects enabled by breaking both space- and time-translation symmetries. Our findings, supported by both experimental data and theoretical models, including full-wave simulations, show that our continuously modulated time-varying absorber can indeed exceed relevant absorption limits, such as those based on the Rozanov bound for LTI absorbers. In light of these results, we also note that the possible extension of these physical limits to time-varying systems represents, in our opinion, an important goal for future theoretical work.

The ability to enhance and control electromagnetic absorption — one of the most fundamental electromagnetic processes — is critically important in a variety of practical scenarios, from stealth technology to energy harvesting and sensing. Looking forward, the principles demonstrated here could pave the way for the development of a new generation of ultrathin, broadband, highly efficient, active electromagnetic absorbers, especially at RF and microwave frequencies, but also potentially in the optical frequency range by using emerging material platforms that support sufficiently fast and deep modulations \cite{hayran2022homega, tirole2023double, harwood2024super}. 
More broadly, by demonstrating the strong potential of temporal degrees of freedom to control waves beyond the fundamental limits of LTI systems, we believe our findings point to new exciting opportunities across various areas of science and technology underpinned by wave physics.

\section*{{Methods}}

\bmhead{Transmission-line circuit design}
We designed the transmission-line circuit to replicate the behavior of a normally incident electromagnetic wave upon a temporally modulated lossy, non-magnetic, dielectric layer followed by a 0.6 m spacer medium, finally backed by a perfect electric conductor (PEC). The circuit is composed of a time varying resistor, a capacitor, and a shorted transmission line (a detailed illustration of the circuit is shown in Fig. \ref{fig2}(A)). The time varying resistor is implemented through the use of a microwave switch with a single-pole-double-throw (SPDT) configuration. The switch alternates between a high resistance and low resistance and is controlled using a square wave control signal (as shown in Fig. \ref{fig3}(A)) oscillating between 0 and 2 V and is biased by applying a positive and negative DC potential of 5 V. The values of resistance the switch alternates between are determined by the following equation: 
\begin{equation}
    R_{\substack{\text{low} \\ \text{high}}} = \frac{R_{0}}{1 \pm M_{\text{mod}}}
    \label{eq10}
\end{equation}
We choose \(M_{\text{mod}}\) = 0.99 to find the values of the required resistances. These are then rounded to the closest available commercial values.
We use the capacitance and inductance per unit length of the transmission line to determine the wavenumber of the transmission line, to then find the correct electrical length of the shorted transmission line \cite{pozar2012microwave}.
\bmhead{Numerical Simulations} 
Both the SI and TSI absorbers, implemented as transmission-line circuits, were simulated with CST Microwave Studio 2022 using the CST EM-Circuit Co-Simulation Transient solver. More details on this full-wave simulation set-up are available in Supplementary Section 5. Simulations of the idealized structures were instead conducted using Lumerical FDTD Solutions. The time-varying absorbing element is modeled as a resistive sheet with a time-varying conductivity, implemented via the Flexible Material Plugin framework. The simulation mesh size is set to 3.26334 mm, and the time step is fixed at 7.62011 ps. The thickness of the resistive sheet is set to equal one mesh size. Metal boundary condition is applied to serve as the reflective PEC surface of the absorber.

\bmhead{Fabrication}
The transmission-line circuits were fabricated on Rogers 4350B (\( \varepsilon_{r} = 3.48 \pm 0.05, \delta = 0.0037 \)) of thickness 0.51 mm and copper layer thickness 35 \( \mu \)m (1 oz. copper). To construct the tranmission-line absorber, we utilized several surface-mount components including a MiniCircuits single-pole-double-throw (SPDT) microwave switch (part no. M3SW-2-50DRA+), a Murata 47 pF capacitor (part no. GRM1885C2E470JW07D), a Vishay 3600 ohm(s) resistor (part no. TNPW04023K60BEED) designated as \( R_{\text{high}} \), TE connectivity 18 ohm(s) resistor (part no. CPF0402B18RE1) designated as \( R_{\text{low}} \). 
For the biasing circuit of the microwave switch, the following components were used: two Kyocera AVX 10 pF capacitors (part no. KGM15ACG2A100GT), two Koa Speer Electronics 11.5 ohm(s) resistors (part no. RK73H1JTTD11R5F), one Koa Speer Electronics 100 ohm(s) resistor (part no. RK73H1JTTD1000F) and two ON semiconductor Zener Diodes with \( V_{z} \) = 5.6 V (part no. MMSZ4690T1G). The circuit assembly was carried out in-house.

\bmhead{Experiments}
The measurement set-up, illustrated in detail in Fig. \ref{fig2}(E), utilizes a Tektronix-AWG520 arbitrary waveform generator (AWG) for generating both the input and control signals. The input signal is defined by:

\begin{equation}
    V_{\text{in}}(t) = \frac{\text{cos}(\omega_{\text{2}}t) - 1}{\omega_{\text{2}} t} - \frac{\omega_{\text{1}}}{\omega_{\text{2}}} \frac{\text{cos}(\omega_{\text{1}}t) - 1}{\omega_{\text{1}} t},
    \label{eq9}
\end{equation}

\noindent where \( \omega_{\text{1}}/2\pi = 8 \) MHz and \( \omega_{\text{2}}/2\pi = 80 \) MHz. This signal is programmed in the AWG using a clock of 1 GHz and time duration 0.5 \( \mu \)s.
A MiniCircuits Directional Coupler (part no. ZMDC-10-1+) was employed to isolate the input and reflected signals, with the reflected signal directed to an Agilent-8563E spectrum analyzer through the coupled port for frequency spectrum analysis. The spectrum analyzer is set to a central frequency of \( \omega_{\text{center}}/2\pi = 252.5 \) MHz and its frequency span is \( \omega_{\text{span}}/2\pi = 495 \) MHz. The measurement is interrupted at 500 MHz since it is the highest operational frequency of the directional coupler and all the harmonics, shown in the inset of Fig. \ref{fig3}(D), have subsided, as confirmed by theoretical/numerical calculations. 
An Agilent-E3630A triple output power supply was used to provide a ±5V DC bias to the microwave switch. To ensure a fair comparison between the input and reflected signals, an SMA open circuit replaced the transmission-line absorber when measuring the input frequency spectrum. This substitution allows the measurement of the frequency spectrum through the coupled port of the directional coupler, ensuring that both the input and reflected signals were subject to the same coupling loss.

\bmhead{Data Analysis}
Our data analysis was focused on the absorption efficiency \( \eta_{\text{abs}} \), as defined in the main text. For each modulation frequency, we obtained 10 reflection spectrum samples for both the reflected and incident signals, \( \Tilde{V}_{\text{r,t}} \) and \( \Tilde{V}_{\text{i,t}} \). To account for the full experimental variability of the absorption efficiency \( \eta_{\text{abs}} \), we calculated its maximum and minimum values based on the extreme values of \( U_{\text{ref}} \) and \( U_{\text{in}} \). These values are determined by:
\begin{align}
\eta_{\text{abs, max}} &= 1 - \frac{\min(U_{\text{ref}})}{\max(U_{\text{in}})}, \\
\eta_{\text{abs, min}} &= 1 - \frac{\max(U_{\text{ref}})}{\min(U_{\text{in}})}.
\end{align}
The error bars in Fig. \ref{fig3}(C) indicate the entire range of \( \eta_{\text{abs}} \), from \( \eta_{\text{abs, min}} \) to \( \eta_{\text{abs, max}} \).

In addition to the range error bars, the error propagation method was used to estimate the uncertainty in \( \eta_{\text{abs}} \) due to the standard deviations of \( U_{\text{ref}} \) and \( U_{\text{inc}} \). The propagated error \( \Delta \eta_{\text{abs}} \) can be calculated in our case as:

\begin{equation}
    \Delta \eta_{\text{abs}} = \sqrt{\left(-\frac{1}{\bar{U}_{\text{in}}} S_{U_{\text{ref}}}\right)^2 + \left(\frac{\bar{U}_{\text{ref}}}{\bar{U}_{\text{inc}}^2} S_{U_{\text{in}}}\right)^2},
\end{equation}

\noindent where \( \bar{U}_{\text{ref}} \) and \( \bar{U}_{\text{in}} \) are the mean values of \( U_{\text{ref}} \) and \( U_{\text{in}} \), respectively, and \( S_{U_{\text{ref}}} \) and \( S_{U_{\text{in}}} \) are their standard deviations. This uncertainty is visually represented in our figures as a color-shaded strip around the mean \( \eta_{\text{abs}} \) value. This dual approach in representing error—using both range error bars and shaded areas for error propagation—provides a thorough depiction of the variability and precision in our \( \eta_{\text{abs}} \) measurements.

\backmatter



\section*{Declarations}

\bmhead{Funding}
Air Force Office of Scientific Research with Grant No. FA9550-22- 1-0204; Office of Naval Research with Grant No. N00014-22-1-2486.

\bmhead{Conflict of interest/Competing interests}
The authors declare no competing interests.

\bmhead{Availability of data and materials}
Authors confirm that all relevant data are included in the paper and/or its Supplementary Information files.

\bmhead{Code availability}
The code used to produce these results is available upon reasonable request to the corresponding author.

\bmhead{Authors' contributions} 
F.M. conceived and supervised the study. M.C. designed and assembled the fabricated structure, designed and built the experimental setup, performed numerical simulations and measurements, and analyzed the experimental data. Z.H. developed the theory, derived the absorption bound, performed theoretical calculations and numerical simulations, analyzed the experimental data, and led the writing of the manuscript with contributions from all authors. All authors participated in the conceptual development, discussion of the results, and the writing of the manuscript.

\bibliography{sn-bibliography}

\end{document}